\documentclass[twoside,fleqn]{myActaStyle}
\usepackage{times,cite}

\usepackage{lscape}             % if you need landscape enviroinment
\usepackage{graphicx,epsfig}    % if you need to include figures in PostScript
\usepackage[tbtags]{amsmath}    % if you need more math

\def\authorheading{F.~Kleefeld}

\def\shorttitle{Coulomb-Scattering and $\eta$-$\eta^\prime$ Mixing Angle}

\begin{document}
%% ---------------------------------------------------

%%%  page range, first and last page
\pagerange{1}{8}

%%% paper title
\title{%
COULOMB-SCATTERING AND $\eta$-$\eta^\prime$ MIXING ANGLE
}

%%% author(s) and address(es)
\author{%  author(s)
Frieder~Kleefeld\;\email{kleefeld@cfif.ist.utl.pt, URL: http://cfif.ist.utl.pt/$\sim$kleefeld/}
}
{%  address(es)
Centro de F\'{\i}sica das Interac\c{c}\~{o}es Fundamentais (CFIF), Instituto Superior T\'{e}cnico,\\ Av.\ Rovisco Pais, 1049-001 Lisboa, Portugal
}

%%% Date of submission
\day{October 4, 2005}

%%% abstract of the paper
\abstract{%
The fascinating physics underlying $\eta$ and $\eta^\prime$ mesons can be studied theoretically and experimentally in various contexts. In this presentation we want to turn our attention to two important uncorrelated aspects of this vivid research field which provide yet unexpected challenges or surprises. First we discuss open questions in the theoretical treatment of Coulomb-interaction in the context of reaction processes like $p p \rightarrow p p \eta$. Then we review $\eta$-$\eta^\prime$ and $\sigma(600)$-$f_0(980)$ mixing in the $U(3)\times U(3)$ Linear Sigma Model and extract information on $\eta$-$\eta^\prime$ mixing and the $K^\ast_0(800)$ resonance from meson-meson scattering.}

%%% PASC numbers of your article
% 13.40.-f = Electromagnetic processes and properties
% 13.75.Lb = Meson-meson interactions
% 13.75.-n = Hadron-induced low- and intermediate-energy reactions 
%            and scattering (energy(less-than-or-equal-to)10 GeV)
% 14.40.Aq = Pi, K, and eta mesons

% \setcounter{section}{1}\setcounter{equation}{0}
\section{Coulomb-Scattering and $pp\rightarrow pp X$ with $X=\pi^0,\;\eta,\;\eta^\prime,\;\ldots$}
Due to their large threshold enhancement (TE) over phasespace \cite{Moskal:2002jm,Moskal:2003gt} strong attention has been devoted to the cross sections of $pp\rightarrow pp\eta$ and $pp\rightarrow pp\eta^\prime$. Despite early theoretical claims that the TE is to be understood by final state interactions (FSI) the conclusive quantitative theoretical explanation of this TE is still awaiting. As long as there is lacking the cross section measurement of ``Coulomb-free'' reaction channels $pn\rightarrow pn\eta$ and $pn\rightarrow pn\eta^\prime$ at excess energies where TE is seen in $pp\rightarrow ppX$ with $X=\eta,\eta^\prime$, it is difficult to say --- lacking also a satisfactory quantitative theoretical description of Coulomb-interactions (CI) in FSI ---, whether the seen TE is solely due to short-ranged strong interactions or whether there is some component of the TE also due to infinite-ranged CI. Without settling these questions it will be therefore close to impossible to learn something quantitatively about the physics underlying $\eta$ and $\eta^\prime$ mesons on the basis of these reaction processes. 
Before entering specific open questions with respect to CI in the context of $pp\rightarrow ppX$ with $X=\eta,\eta^\prime$ we shortly want to recall the respective present theoretical situation concerning FSI: Present theoretical approaches to the treatment of initial state interactions~(ISI) and FSI are typically based on a ``two-potential formalism'' (TPF), where the overall interaction potential $V$ decomposes according to $V=V_S+V_L$ into a sum of a short-ranged ($V_S$) and a long-ranged ($V_L$) interaction potential, while $V_S$ leads to production of meson $X$ and $V_L$ dominiates ISI and FSI.\footnote{Accordingly the unitary scattering matrix is assumed to decompose as \cite{Kleefeld:2001gz} $S = (S^{-1})^+ =\exp(2\mathrm{i} (\Delta_S + \Delta_L)) = \exp(\Delta_L \frac{\delta}{\delta\Delta_S}) \exp(2\mathrm{i} \Delta_S)\exp(-\Delta_L \frac{\delta}{\delta\Delta_S})\equiv T_{FSI}\exp(2\mathrm{i} \Delta_S)T_{ISI}$. 
Here $\Delta_S$ and $\Delta_L$ are Hermitian symmetric phaseshift ``matrices'' constrained by the (functional) commutator $[\delta/(\delta\Delta_S),\Delta_S]=1$. The ``Enhancement Factors'' (EFs) $T_{FSI}$ and $T_{ISI}$ {\em are not independent} due to $T_{ISI}=T^{-1}_{FSI}$, while the unitarity condition $T^{-1}_{FSI}=T^+_{FSI}$ implies $[\delta/(\delta\Delta_S),\Delta_L]=0$ showing that EFs {\em depend on the short ranged interactions} \cite{Baru:2000vd}. 
The TPF goes back to work by \cite{Brueckner:1951} Brueckner, Chew, Hart (1951), Watson (1951,1952), Gell-Mann, Goldberger (1953), Migdal (1955), Fermi (1955). The TPF for ISI/FSI in $2\rightarrow 3$ processes for finite range interactions was worked out already in detail as early as 1968 \cite{Aitchison:1969tq}.} Despite its simple philosophy the technical implementation of the TPF gets rather cumbersome and requires particular care, when $V_L$ happens to be of infinite range like in the case of CI. The technical challenge related e.g.\ to CI has been a combination of several aspects. First of all one is in the need to be able to describe theoretically elastic scattering (ES) of any two charged particles occuring in either the initial or the final state.\footnote{The determination of required wave functions or on-shell Green's functions (and related integrals) for non-relativistic two-particle Coulomb ES has a long tradition (e.g.\ \cite{Gordon:1928,VanHaeringen:1985tp,Ahmed:2003}) and has been known to be quite a rewarding technical task which is performed for on-shell Green's functions most conveniently in momentum space \cite{Okubo:1960b,Ford:1964}.} The second level of complication has been to put either incoming or outgoing particles of the non-relativistic Coulomb-scattering (CS) problem off the mass shell and to calculate corresponding ``half-shell'' \cite{Kowalski:1965} Green's functions \cite{Ford:1964,Kok:1982,Talukdar:1984,VanHaeringen:1985tp,Katsogiannis:1994} or off-shell Jost functions \cite{Kok:1982,Talukdar:1984,VanHaeringen:1985tp,Katsogiannis:1994} for the non-relativistic CS problem. The surprising observation of related studies is \cite{Ford:1964,Chen:1971,Kok:1981,Kok:1982,VanHaeringen:1985tp,Alston:1988,Katsogiannis:1994} that the half-shell Green's function of non-relativistic CS is not related to the respective on-shell Green's function by a smooth limiting procedure.\footnote{Similarly, half-shell Green's functions of Hulth\'{e}n-like potentials \cite{Ford:1964,Kok:1982} display a rapid change at the on-shell limit.} This implies in particular that the on-shell unitarity relation for the non-relativistic Coulomb Green's function is not well defined \cite{Chen:1971}. The third challenge has been then to be able to describe half-shell non-relativistic CS in the presence of short-ranged nuclear interactions \cite{Dolinskii:1966,Kok:1981,Katsogiannis:1994}. As half-shell Green's functions for $2\rightarrow 2$ scattering processes determine merely EFs for interacting two-particle subsystems in the initial or final state, there arises the final --- yet unresolved --- challenge to determine EFs for $2\rightarrow 3$ production processes for the case of infinite ranged FSI involving three scattering particles. The most popular, yet --- as we will argue below --- incorrect assumption in the spirit of the distorted wave impulse approximation for short ranged interaction potentials is that even infinite range Coulomb FSI corrections are factorizable. In the context of quasi-free processes with two or three charged particles in the final state factorization of Coulomb FSI has been promoted e.g.\ in Refs. \cite{Bajzer:1987}. The authors of Ref.\ \cite{Moalem:1995} tried to achieve factorization by performing an on-shell approximation in the three particle final state. The authors of Ref.\ \cite{Alt:2004fi} seem to make the strongest case for factorization by performing a so-called ``Coulomb Fourier Transform''. Unfortunately their calculations require regularization and are therefore in the end regularization scheme dependent. As pointed out in Ref.\ \cite{Mukhamedzhanov:2005xc} the regularization of CI for systems of two charged particles by a screening technique \cite{Alt:1978yd} being based on foregoing ideas \cite{Dollard:1968} is ``successfully implemented''. Yet also here it is quite rewarding a challenge \cite{Deltuva:2005wx} to remove the inherent regulator dependence. 
Parallely to the aforementioned theoretical machinery with all its difficulties there appeared \footnote{Unaware of the existence of a worked out formalism of 1968 \cite{Aitchison:1969tq} for ISI/FSI in $2\rightarrow 3$ processes for finite-ranged interactions theoreticians involved in nucleon-induced threshold meson production tried around 1997  \cite{Faldt:1997jm,Kleefeld:2001gz,Baru:2000vd} to re-develop a formalism for the theoretical description of ISI/FSI, yet without managing to handle infinite-ranged CI quantitatively.} in 2003 a surprisingly simple and seemingly successful formalism by A.\ Deloff \cite{Deloff:2003te} who not only constructed final state EFs on the basis of 3-body wavefunctions, yet also invoked in such 3-body EFs gross features of CI. 
His method applied to $pp\rightarrow pp\eta$ allowed him \cite{Deloff:2003te} --- assuming still a constant meson production amplitude --- to obtain already at leading order of the partial wave expansion a quite impressive near to quantitative description of the Dalitz-plot $\frac{\mathrm{d}^2\sigma}{\mathrm{d}s_{pp} \mathrm{d}s_{p\eta}}$ close to threshold. In normalizing the results at small values of the outgoing proton invariant mass square $s_{pp}$ to experimental data he observed that theoretical predictions are too small compared to experiment at higher values of $s_{pp}$, where the outgoing $\eta$-meson is approximately at rest. Although this deviation of theory from experiment for $s_{pp}$ large is presently believed to be due to an insufficient theoretical description of short-ranged strong interactions, we want to point here out that that discrepancy between theory and experiment may be also {\it due to an insufficient treatment of Coulomb FSI} in $pp\rightarrow pp\eta$. A first way to understand this conjecture is to recall that the radius of convergence of the partial wave expansion in the very vincinity of a Coulomb-singularity is very small. This implies for a reaction like $pp\rightarrow pp\eta$ that already at energies very close to threshold a very large number of partial waves even of large $pp$ orbital momenta have to be taken into account \cite{Ahmed:2003,Deltuva:2005wx} to obtain an accurate description of the reaction amplitude in the complete Dalitz plot. 
Another way of understanding the conjecture is achieved by looking at $pp$ ES, where the total cross section $\sigma$ is naively believed to diverge due to a non-integrable singularity of the Rutherford differential cross section $\frac{\mathrm{d}\sigma}{\mathrm{d}\Omega}\simeq \frac{\alpha^2}{(2 m v^2 )^2}\frac{1}{\sin^4(\theta/2)}$ at $\theta\rightarrow 0$. This paradox is resolved \cite{Baryshevskii:2004,Mukhamedzhanov:2005xc} by taking into account in calculations for $\sigma$  not only the divergent asymptotic part of outgoing pp-wavefunction, yet the {\em finite} full wavefunction $\psi_{\vec{k}}(r)$ of the ES problem containing also the divergence contained in the incoming $pp$-system. 
The \textit{finite} total cross section is then calculated on the basis of the current density $\vec{j} = \frac{1}{2\mathrm{i}m} (\psi^\ast_{\vec{k}}(r) \vec{\nabla} \psi_{\vec{k}}(r) - \psi_{\vec{k}}(r) \vec{\nabla} \psi^\ast_{\vec{k}}(r))$ (with $\vec{j}\rightarrow \frac{\vec{k}}{m} \equiv \vec{j}_0$ for $\vec{k}\cdot\vec{r}\rightarrow -\infty$ and $\vec{j}  \rightarrow j_e \frac{\vec{k}}{k} + j_{sc} \frac{\vec{r}}{r}$ for $r\rightarrow\infty$) as  \cite{Baryshevskii:2004} $\sigma = \int \mathrm{d}\Omega \,r^2 \frac{j_{sc}}{j_0}  = \frac{2\pi r}{k} \, \xi^2  I_-(\xi)$ with $I_-(\xi)\equiv \exp(\pi \xi) \int^\infty_0dz \, |U_1(1-\mathrm{i}\xi,1,\mathrm{i}z)|^2$ and $\xi\equiv\frac{\alpha}{v}$. $\sigma$ shows here up to be not only energy- yet also \textit{volume-dependent}, and diverges only for an infinite reaction volume, i.e.\ $r\rightarrow \infty$!~\footnote{Also Ref.\ \cite{Pineda:2004mx} observed that the definition of the proton charge radius depends on some reaction dependent length.} The correct calculation of $\sigma$ required --- contrary to what is assumed in the TPF --- a \textit{knowledge of the $pp$-wavefunction in the reaction point}. 
These observations on $pp$ ES get relevant for the discussion of FSI/ISI in $pp\rightarrow ppX$ with $X=\eta,\eta^\prime$ when taking into account most recent conclusions of Ref.\ \cite{Mukhamedzhanov:2005xc} studying $3\rightarrow 3$ scattering of three charged particles: {\em ``\ldots If any of the particles is neutral, then the resulting asymptotic solution becomes the plane wave for the neutral particle and the exact two-body scattering wavefunction for the charged pair \ldots''}. 
This implies that due to the neutrality of $X=\eta,\eta^\prime$ the protons in $pp\rightarrow ppX$ should show in particular for the meson $X$ at rest (i.e. $s_{pp}$ large) some features observed also in $pp$ ES. By the foregoing considerations one might understand now, why \mbox{$\sigma(pp\rightarrow ppX)$} with $X=\eta,\eta^\prime$ is --- {\em despite} the influence of CI --- finite at all and why the experimental Dalitz-plot $\frac{\mathrm{d}^2\sigma}{\mathrm{d}s_{pp} \mathrm{d}s_{p\eta}}$ is showing --- eventually {\em due} to CI ---  some enhancement beyond naive theoretical expectations for large $s_{pp}$. Further we expect some {\em reaction-volume dependence} of total cross sections in the presence of CI, which might be different for $pp\rightarrow pp\eta$ and $pp\rightarrow pp\eta^\prime$ and experimentally explored by correlation functions like the one proposed by Pawe\l~Klaja \cite{Klaja:2005a} during the Eta'05 workshop in Cracow. We are left with the puzzling observation that angular distributions in $pp\rightarrow pp\eta$ seem to be flat \cite{Moskal:2003gt}, contrary to what is observed in $pp$ ES. During the same Eta'05 workshop it has been also argued on the basis of Ref.\ \cite{Gasparyan:2005fk} that CI in $pp\rightarrow pp\eta$ are fully understood. We cannot share this belief due to the fact, that the dispersive method displayed in Ref.\ \cite{Gasparyan:2005fk} being based on analyticity assumptions to be yet justified is used only to calculate effective range parameters for systems of charged particles with different mass, and NOT complete $2\rightarrow 3$ production cross sections for the more pathologic situation of two charged initial and final state particles of {\em equal mass} like it is the case for $pp\rightarrow ppX$ with $X=\eta,\eta^\prime$. Also the method does NOT provide any solution to what has been summarized in Ref.\ \cite{Kadyrov:2005xm} as follows: {\em ``$\ldots$ For the charged particles with the long-range Coulomb interaction the theory has faced apparently insurmountable difficulties. The problem is that the Faddeev equations are not compact in the presence of Coulomb interactions. $\ldots$''} 
\section{$\eta$-$\eta^\prime$ Mixing Angle}
\subsection{``Traditional'' $U(3)\times U(3)$ Linear $\sigma$ Model Approach to $\eta\,$-$\,\eta^\prime$ Mixing} \label{seclinsig1}
We want to recall here some ``traditional'' one-mixing-angle approach to $\eta\,$-$\,\eta^\prime$ and $\sigma$-$f_0$ mixing in the context of the $U(3)\times U(3)$ Linear $\sigma$ Model (L$\sigma$M) (For $\pi^0$-$\eta$ mixing in the $U(2)\times U(2)$ L$\sigma$M see 't~Hooft (1986)\cite{Levy:1967a}). 
For later convenience we define $\Sigma_\pm (x) \equiv S(x) \pm i P(x)$ and the following $U(3)\times U(3)$ scalar and pseudoscalar meson field matrices
\begin{equation} S = \left( 
\begin{array}{ccc} \sigma_{u\bar{u}} & a^+_0 & \kappa^+ \\[1mm]
a^-_0 & \sigma_{d\bar{d}} & \kappa^0 \\[1mm]
\kappa^- & \bar{\kappa}^0 & \sigma_{s\bar{s}}
\end{array}\right) \; , \quad P = \left( 
\begin{array}{ccc} \eta_{u\bar{u}} & \pi^+ & K^+ \\[1mm]
\pi^- & \eta_{d\bar{d}} & K^0 \\[1mm]
K^- & \bar{K}^0 & \eta_{s\bar{s}}
\end{array}\right) \; , 
\end{equation}
and $\sigma_{n\bar{n}}\equiv (\sigma_{u\bar{u}} + \sigma_{d\bar{d}})/\sqrt{2}$, $\sigma_{3}\equiv (\sigma_{u\bar{u}} - \sigma_{d\bar{d}})/\sqrt{2}\simeq a^0_0$, and $\eta_{n\bar{n}}\equiv (\eta_{u\bar{u}} + \eta_{d\bar{d}})/\sqrt{2}$, $\eta_{3}\equiv (\eta_{u\bar{u}} - \eta_{d\bar{d}})/\sqrt{2}\simeq \pi^0$.
The Lagrangean of the $U(3)\times U(3)$ L$\sigma$M before spontaneous symmetry breaking --- for simplicity without (axial) vector mesons --- is given by \cite{Levy:1967a,Delbourgo:1998kg,Scadron:2006mq}:
\begin{eqnarray}
{\cal L}  & = & \frac{1}{2} \, \mbox{tr} [(\partial_\mu \Sigma_+)(\partial^\mu \Sigma_-)] - \frac{1}{2} \, \mu^2 \, \mbox{tr} [ \Sigma_+ \Sigma_-]  - \frac{\lambda}{2} \, \mbox{tr} [ \Sigma_+ \Sigma_- \Sigma_+ \Sigma_-] \nonumber \\
 & & - \frac{\lambda^\prime}{4} \, \Big( \mbox{tr} [ \Sigma_+ \Sigma_-]\Big)^2 + \frac{\beta}{2} \Big( \mbox{det} [ \Sigma_+] + \mbox{det} [ \Sigma_-]\Big) + \mbox{tr} [ C S] \; . \label{eqlsm1}
\end{eqnarray}
Eq.\ (\ref{eqlsm1}) containing direct chiral symmetry breaking due to the term $\mbox{tr} [C S]$ with $C$ being a constant diagonal $3\times 3$-matrix and containing $U_A(1)$-symmetry breaking due to the 't~Hooft determinant term \cite{Levy:1967a} proportional to $\beta$ being --- as we shall see below --- responsible for $\eta$-$\eta^\prime$ mixing is stabilized by performing spontaneous symmetry breaking, i.e.\ by performing a (here approximately isospin symmetric) shift $S\rightarrow S-D$ with $D\simeq\mbox{diag} (a,a,b)$ such that ${\cal L}_1 = \mbox{tr} [S (2\lambda D^3 + \beta D^2 + (\mu^2 + \lambda^\prime \, \mbox{tr}[D^2]-\beta \, \mbox{tr} D )D + \frac{\beta}{2} ((\mbox{tr} D)^2 -\mbox{tr} [D^2] ) + C)]$
vanishes. $f_\pi = \sqrt{2}\, a = 92.4$~MeV and $f_K=(a+b)/\sqrt{2}$ are the pion and kaon decay constants. The (isospin symmetric) mass Lagrangean of the spontaneously broken $U(3)\times U(3)$ L$\sigma$M is
\begin{eqnarray} {\cal L}_2 & = &  - \frac{1}{2} \,\Big( m^2_{a_0} \, (2 \, a^+_0 a^-_0 + \sigma^2_3 )  + 2 \, m^2_\kappa\,(\kappa^+ \kappa^- + \kappa^0 \bar{\kappa}^0) + m^2_{\sigma_{n\bar{n}}} \sigma^2_{n\bar{n}} + m^2_{\sigma_{s\bar{s}}} \sigma^2_{s\bar{s}}\nonumber \\
 & & \quad\;\; + m^2_\pi \, (2 \, \pi^+ \pi^- + \eta^2_3 )\; + 2\,m^2_K \, (K^+ K^- + K^0 \bar{K}^0) + m^2_{\eta_{n\bar{n}}} \eta^2_{n\bar{n}} \,+ m^2_{\eta_{s\bar{s}}} \eta^2_{s\bar{s}} \nonumber \\
 & & \quad\;\;  + 2 \sqrt{2}\,a\, (\beta + 2\lambda^\prime b) \,\sigma_{n\bar{n}} \sigma_{s\bar{s}} - 2 \sqrt{2}\,a\, \beta \,\eta_{n\bar{n}} \eta_{s\bar{s}} \Big) \; , \label{eqlsm2} \end{eqnarray} 
with $\bar{\mu}^2 \equiv \mu^2 + \lambda^\prime (2a^2 + b^2)$ and 
$m^2_{a_0} = \bar{\mu}^2 + 6\lambda a^2 - \beta b$, 
$m^2_\kappa = \bar{\mu}^2 + 2\lambda (a^2+b^2+ab) - \beta a$, 
$m^2_\pi = \bar{\mu}^2 + 2\lambda a^2 + \beta b$,
$m^2_K= \bar{\mu}^2 + 2\lambda (a^2+b^2-ab) + \beta a$,
$m^2_{\sigma_{n\bar{n}}}= \bar{\mu}^2 + (6\,\lambda + 4 \lambda^\prime ) a^2 + \beta b$, 
$m^2_{\sigma_{s\bar{s}}} = \bar{\mu}^2 + (6\,\lambda + 2 \lambda^\prime ) b^2$,
$m^2_{\eta_{n\bar{n}}} =\bar{\mu}^2 + 2\,\lambda a^2 - \beta b$,
$m^2_{\eta_{s\bar{s}}} = \bar{\mu}^2 + 2\,\lambda b^2$. Eq.\ (\ref{eqlsm2}) is diagonalized by $\sigma(600) = \sigma_{n\bar{n}} \cos \phi_S - \sigma_{s\bar{s}} \sin \phi_S$, $f_0(980) = \sigma_{n\bar{n}} \sin \phi_S + \sigma_{s\bar{s}} \cos \phi_S$, and  $\eta(548) = \eta_{n\bar{n}} \cos \phi_P - \eta_{s\bar{s}} \sin \phi_P$, $\eta^\prime(958) = \eta_{n\bar{n}} \sin \phi_P + \eta_{s\bar{s}} \cos \phi_P$. The observation of an empirical ``Equal Splitting Law'' $m^2_{\sigma_{n\bar{n}}} - m^2_\pi \simeq m^2_{a_0} - m^2_{\eta_{n\bar{n}}}$ yielding $\lambda^\prime a^2 \simeq 0$ and therefore $\lambda^\prime \simeq 0$ for $a\not=0$ allowed Scadron \textit{et al.} \cite{Klabucar:2001gr,Delbourgo:1998kg} to determine $m_{\eta_{n\bar{n}}}\simeq 757.9$~MeV from $m^2_{\sigma_{n\bar{n}}} - m^2_\pi \simeq (2\hat{m})^2$ assuming $m_{a_0} \simeq 984.8$~MeV and $\hat{m}= \sqrt{\lambda a^2}$ being the nonstrange quark-mass obtained aproximately as one third of the nucleon mass, i.e.\ $\hat{m} \simeq m_N / 3 \simeq 315\,\mbox{MeV}/3$. 
This result led \cite{Klabucar:2001gr} on the basis of $m^2_{\eta_{n\bar{n}}} = m^2_\eta \cos^2\phi_P + m^2_{\eta^\prime} \sin^2\phi_P$ and $m^2_{\sigma_{n\bar{n}}} = m^2_\sigma \cos^2\phi_S + m^2_{f_0} \sin^2\phi_S$ to pseudoscalar and scalar mixing angles in the nonstrange-strange (``ideal'') basis $\phi_P=\arctan ((m^2_{\eta_{n\bar{n}}} - m^2_\eta)/(m^2_{\eta^\prime} - m^2_{\eta_{n\bar{n}}}))^{1/2} \simeq  41.84^\circ$ \cite{Klabucar:2001gr} and $\phi_S=\arctan ((m^2_{\sigma_{n\bar{n}}} - m^2_{\sigma})/(m^2_{f_0} - m^2_{\sigma_{n\bar{n}}}))^{1/2}  \simeq \pm 18^\circ$ \cite{Klabucar:2001gr,Kleefeld:2001ds}. The obtained value for $\phi_P$ is very consistent with newer \cite{Kroll:2005sd,Escribano:2005qq,Scadron:2006mq} and compatible with older \cite{Bramon:1997mf} experimental and theoretical findings. Most recent KLOE experimental data suggest \cite{Kroll:2005sd} e.g.\ $\phi_P= \, (41.2\pm 1.1)^\circ$.  
\subsection{$(K\pi,K\eta,K\eta^\prime,\ldots)\rightarrow (K\pi,K\eta,K\eta^\prime,\ldots)$ Scattering and the $\eta$-$\eta^\prime$ Mixing Angle} \label{seccoupled1}
\begin{figure}[t]
\begin{minipage}[t]{0.32\linewidth}
\begin{center}
\includegraphics[width=1.7in]{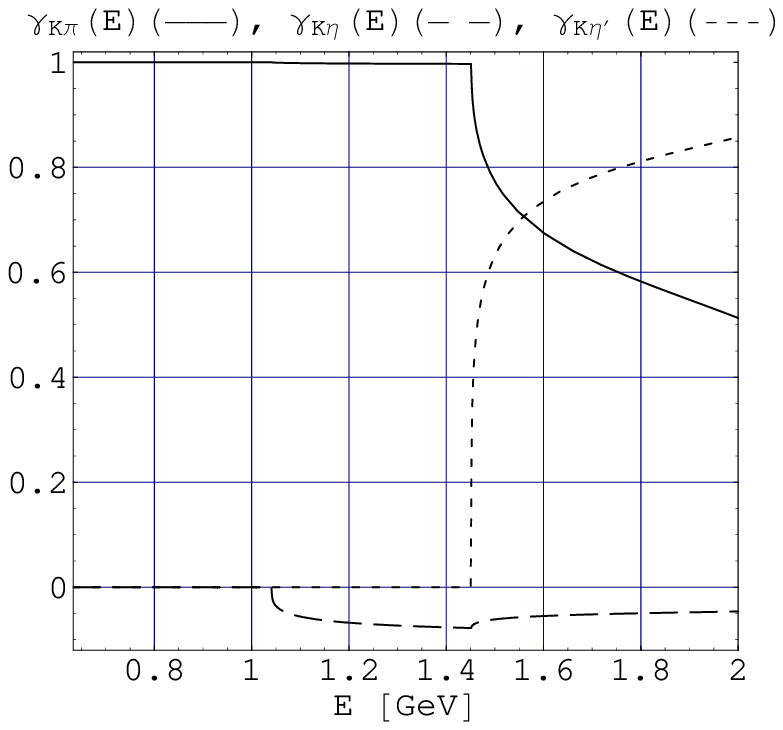} \\
\end{center}
\caption{$E$-dependence of $\gamma_{K\pi} (E)$ (solid line), $\gamma_{K\eta} (E)$ (dashed line), $\gamma_{K\eta^\prime} (E)$ (dotted~line).
}
 \label{fig1}
\end{minipage}%
\hspace{0.02\textwidth}%
\begin{minipage}[t]{0.32\linewidth}
\begin{center}
\includegraphics[width=1.7in]{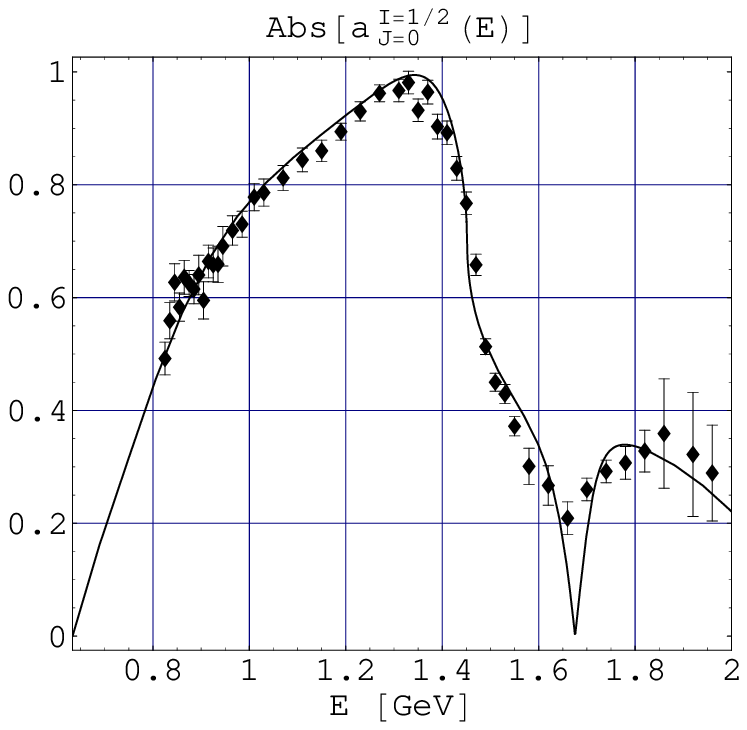} \\
\end{center}
\caption{LASS-data for the modulous of the $K\pi\rightarrow K\pi$ amplitude $|a^{I=1/2}_{J=0}(E)|$ and fit (solid line).
}\label{fig2}
\end{minipage}
\hspace{0.02\textwidth}%
\begin{minipage}[t]{0.32\linewidth}
\begin{center}
\includegraphics[width=1.7in]{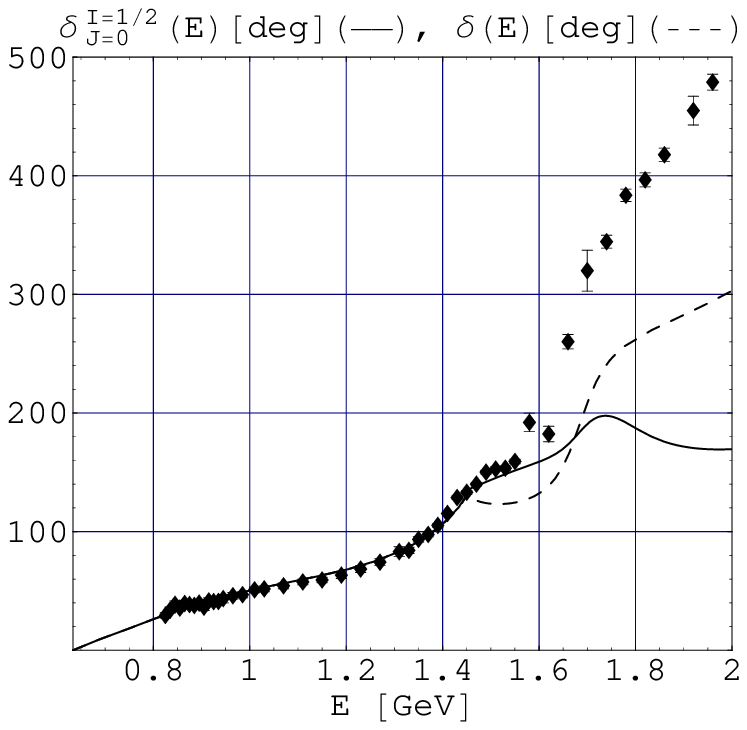} \\
\end{center}
\caption{RSE fit for $\delta^{I=1/2}_{J=0}(E)$ (solid line) and eigenphase $\delta(E)$ (dotted line) versus LASS-data. 
}
 \label{fig3}
\end{minipage}%
\end{figure}

Despite the predictive power of a Lagrangean approach like in Section \ref{seclinsig1} related ``tree-level'' results should be interpreted with great precaution due to ``unitarization effects'' \cite{Truong:1991gv} relating a ``tree-level'' Lagrangean and the (non-perturbative) effective action.\footnote{Unitarization leads for one field in a tentative Lagrangean typically to several distinct poles of the scattering matrix~(S-matrix) \cite{Kleefeld:2003bw,vanBeveren:2002gy}. Therefore one field in a tentative Lagrangean like Eq.\ (\ref{eqlsm1}) typically corresponds to several experimental ``particles'', the pole parameters of which can be very different from the mass parameters of the original tentative Lagrangean, if unitarization effects are large. Eq.\ (\ref{eqlsm1}) provides us with a gold-plated example for large unitarization effects: the $\kappa$-field in Eq.\ (\ref{eqlsm1}) is typically described by a mass parameter of approximately $m_\kappa \simeq 1150$~MeV (e.g.\ T\"ornqvist (1999) \cite{Levy:1967a}), while unitarization then yields at least two known experimental mesons $K^\ast_0(800)$ and $K^\ast_0(1430)$, the pole parameters of which are obviously very different from the Lagrangean parameter $m_\kappa$.} Certainly, a tentative Lagrangean like Eq.\ (\ref{eqlsm1}) may be for some fields contained already very close to the effective action corresponding to specific poles of the scattering matrix. In this ``bootstrapping'' situation unitarization effects are --- up to generation of extra poles --- very small due to an at least approximate cancellation of all non-tree-level diagrams related to these fields. This is why in Eq.\ (\ref{eqlsm1}) resonances like $\sigma(600)$ and $f_0(980)$ are described already close to tree-level, even if we have to be aware that unitarization leads to extra ``particles'' with equal quantum numbers like $f_0(1370)$, $f_0(1500)$, $f_0(1700)$, $\ldots$.  
Hence, we want to understand below, whether the conclusions of Section \ref{seclinsig1} survive within a fully unitarized framework and how ``mixing'' between short-lived ``particles'' like in the $\sigma$-$f_0$ system or much more long-lived ``particles'' like in the $\eta$-$\eta^\prime$ system parametrized by (eventually complex-valued) mixing angles expresses itself as a consequence of unitarization in experimental observables. An instructive way to study unitarization effects and $\eta$-$\eta^\prime$ ``mixing'' is scattering of pseudoscalar mesons with total isospin $I=1/2$ and angular momentum $J=0$, probing directly the resonances like $K^\ast_0(800)$ and $K^\ast_0(1430)$. For energies of our interest it appears (naively) sufficient to take into account the 3 lowest lying thresholds, i.e.\ to consider unitarized scattering of 3 coupled channels $K\pi$, $K\eta$, and $K\eta^\prime$, which will be here described by the so-called ``Resonance Spectrum Expansion'' (RSE) \cite{vanBeveren:2001kf,Kleefeld:2003bw} $\bar{g}(E) = (\lambda^2/a) \sum_n B_n/(E-E_n)$ inspired by the ``Nijmegen Unitarized Meson Model'' \cite{ruppthesis1,VanBeveren:1986ea} (NUMM). 
Within the framework of the RSE we determine the S-matrix of the radial meson-meson-scattering Schr\"odinger equation $(\frac{d^2}{dr^2} + {\cal K}^2) \vec{\psi}(r) = 2\mu_S \,\bar{G} (E)\, \delta(r-a) \vec{\psi}(r)$ with $\vec{\psi}(r)=(\psi_{K\pi}(r),\psi_{K\eta}(r),\psi_{K\eta^\prime}(r))^T$, $2\mu_S=\mathrm{diag}(\mu_{K\pi},\mu_{K\eta},\mu_{K\eta^\prime})$, ${\cal K}=\mathrm{diag}(k_{K\pi},k_{K\eta},k_{K\eta^\prime})$, $M_{ij}\equiv m_i + m_j$, $\mu_{ij}\equiv m_im_j/M_{ij}$, $E=(k^2_{ij}+m^2_i)^{1/2} + (k^2_{ij}+m^2_j)^{1/2}$ and $i,j\in \{K,\pi,\eta,\eta^\prime\}$. The $E$-dependent symmetric coupling matrix $\bar{G}(E)$ of the $\delta$-shell transition potential of range $a$ between the meson-meson scattering continuum and the 1-channel confining quark-antiquark ($q\bar{q}$) system is given by $\bar{G}(E) = \sqrt{2\mu(E)/(2\mu_S)} \;\, \vec{V} \, \bar{g}(E) \, \vec{V}^+ \! \sqrt{2\mu(E)/(2\mu_S)}$ with $\vec{V}=(V_{K\pi},V_{K\eta},V_{K\eta^\prime})^T$, $2\mu(E)=\mathrm{diag}(\mu_{K\pi}(E),\mu_{K\eta}(E),\mu_{K\eta^\prime}(E))$, $\mu_{ij}(E)\equiv (E^4 - (m^2_i - m_j^2)^2)/(2 E^3)$ and $i,j\in \{K,\pi,\eta,\eta^\prime\}$ \cite{Kleefeld:2003bw}. 
The flavour blind meson-quark recoupling constants are taken from \cite{VanBeveren:1986ea} as $V_{K\pi} = 1/\sqrt{16}$, $V_{K\eta} = (\cos\phi_P - \sqrt{2} \, \sin\phi_P)/\sqrt{48}$, $V_{K\eta^\prime} = (\sin\phi_P + \sqrt{2} \, \cos\phi_P)/\sqrt{48}$. As scattering proceeds through {\em one} $q\bar{q}$ channel only, the resulting $3\times 3$ scattering matrix ${\cal S}(E)=1-{\cal P}(E) + {\cal P}(E) \exp(2\mathrm{i}\delta(E))$ is characterized by {\em one} ``eigenphase'' $\delta(E)$ only, while ${\cal P}(E)=\vec{\gamma}(E) \vec{\gamma}(E)^T$ is a rank 1 projector with effective $E$-dependent recoupling constants $\vec{\gamma}(E)=(\gamma_{K\pi}(E),\gamma_{K\eta}(E),\gamma_{K\eta^\prime}(E))$ (see Fig.~\ref{fig1}) which for {\em real} $E$ are real and normalized according to  $\vec{\gamma}(E)\cdot \vec{\gamma}(E)=1$. Up to an overall normalization we obtain $\gamma_{ij}(E)\propto \theta (E-M_{ij}) V_{ij} \sin(a k_{ij}) \sqrt{2\mu_{ij}(E)/(a k_{ij})}$ with $i,j\in \{K,\pi,\eta,\eta^\prime\}$ being obviously {\em independent} of the RSE $\bar{g}(E)$! 
After choosing for the RSE the reasonable ansatz with two confinement bare states and one background term ($E_{-1}=0$), i.e.\ $a\,\bar{g}(E) \simeq \frac{\bar{B}_{-1}}{E} +  \frac{\bar{B}_0}{E-E_0} +  \frac{\bar{B}_1}{E-E_1}$ with $\bar{B}_j\equiv \lambda^2 B_j$ ($j=0,\pm 1$), we perform a fit (see Fig.~\ref{fig2}) of the modulous $|a^{I=1/2}_{J=0}(E)|=|\sin\delta^{I=1/2}_{J=0}(E)|$ of the $I=1/2$ scalar $K\pi\rightarrow K\pi$ amplitude measured at the LASS-spectrometer in 1988 \cite{Aston:1987ir}.\footnote{We are aware of strong unitarity violations of considered LASS data for energies higher than $E\approx 1.5$~GeV which should be seriously taken into account when interpreting our results for $E\ge 1.5$~GeV!} Starting from parameters $a=2.55\;\mbox{GeV}^{-1}$, $\bar{B}_{-1} = -17.49$, $\bar{B}_{0} = 3.5$, $\bar{B}_{1} = 1$, $E_0=1.46\;\mbox{GeV}$, $E_1=1.85\;\mbox{GeV}$ guessed in the elastic region of the data a fit with the help of the {\em Mathematica} fit function \mbox{\sf FindFit} returns the result $a \simeq 2.5497\;\mbox{GeV}^{-1}$, $\bar{B}_{-1} \simeq  -12.7387$, $\bar{B}_{0} \simeq  4.6931$, $\bar{B}_{1} \simeq 3.1380$, $E_0 \simeq 1.5173$~GeV, $E_1 \simeq 1.8178$~GeV. The fit being performed at a mixing angle $\phi_P = 39.40^\circ$ compatible with Ref.\ \cite{Escribano:2005qq} yields a S-matrix pole for the $K^\ast_0(800)$-meson at $(736.10-\mathrm{i}257.83)$~MeV (to be compared to \cite{Bugg:2003kj,Scadron:2006mq}). Fig.~\ref{fig3} shows that $\phi_P = 39.40^\circ$ is still compatible with the experimental nil-result that --- using the words of T\"ornqvist (1995) \cite{vanBeveren:2002gy} --- \textit{``\ldots the $K\eta$ threshold essentially decouples because of the small coupling constant \ldots''}. Had we exactly decoupled the $K\eta$-channel by choosing $V_{K\eta}=0$, then the resulting mixing angle would have been obviously $\phi_P=\arctan (1/\sqrt{2})=35.26^\circ$. Note from Fig.~\ref{fig1} that --- despite constant $\phi_P$ --- the effective couplings $\vec{\gamma}(E)$ vary strongly with $E$! Sections \ref{seclinsig1} and \ref{seccoupled1} are intimately related: on one hand mesons in the NUMM/RSE interact only {\em indirectly} by coupling to confining $q\bar{q}$ channels, on the other hand Eq.\ (\ref{eqlsm1}) is obtained by allowing mesons to couple to $q\bar{q}$ only \cite{Kleefeld:2005hd}.  
%  Fig. 1,   Fig. 1

\begin{ack} This work summarizes our oral contribution at the Eta'05 workshop on ``Production and Decay of $\eta$ and $\eta^\prime$ Mesons'', 15-18.9.2005, Cracow, Poland. We are most grateful to Pawe\l~Moskal, the Research Centre in J\"ulich and the EtaMesonNet for financial support, kind invitation to and hospitality during this workshop! The work has been supported by the
{\em Funda\c{c}\~{a}o para a Ci\^{e}ncia e a Tecnologia} \/(FCT) of the {\em Minist\'{e}rio da Ci\^{e}ncia, Tecnologia e Ensino Superior} \/of Portugal, under Grants no.\ PRAXIS XXI/BPD/20186/99, SFRH/BDP/9480/2002, POCTI/\-FNU/\-49555/\-2002, and POCTI/FP/FNU/50328/2003.
\end{ack}


\begin{thebibliography}{99}
%%%%%%%%%%%%%%%%%%%% Coulomb %%%%%%%%%%%%%%%%%%%%%%%%%%%%%%%%%%
\bibitem{Moskal:2002jm}
\refer{P.~Moskal {\it et al.}}{Prog. Part. Nucl. Phys.}{49}{2002}{1};
%%CITATION = HEP-PH 0208002;%%
% \bibitem{Moskal:2004cm}
P.~Moskal: habilitation thesis, Jagellonian Univ., Cracow, Poland, 2004, ISBN 83-233-1889-1 [hep-ph/0408162];
%%CITATION = HEP-PH 0408162;%%
% \bibitem{Khoukaz:2004si}
\refer{A.~Khoukaz {\it et al.}}{Eur. Phys. J. A}{20}{2004}{345};
%%CITATION = NUCL-EX 0401011;%%
% \bibitem{Przerwa:2005ir}
J.~Przerwa {\it et al.}: hep-ex/0507076;
%%CITATION = HEP-EX 0507076;%%
% \bibitem{Moskal:2002wp}
\refer{P.~Moskal {\it et al.}}{Nucl. Phys. A}{721}{2003}{657}
%%CITATION = NUCL-EX 0212003;%%

\bibitem{Moskal:2003gt}
%``Experimental study of p p eta dynamics in the p p $\to$ p p eta reaction,''
\refer{P.~Moskal {\it et al.}}{Phys. Rev. C}{69}{2004}{025203}
%%CITATION = NUCL-EX 0307005;%%

\bibitem{Kleefeld:2001gz}
%``ISI and FSI in N N $\to$ N N X reactions close to threshold,''
\refer{F.~Kleefeld}{Schriften des FZ J\"ulich, Series Matter and Material}{11}{2002}{51} [nucl-th/0108064]
%%CITATION = NUCL-TH 0108064;%%

\bibitem{Baru:2000vd}
%``FSI effects in meson production in N N collisions,''
\refer{V.~Baru {\it et al.}}{Acta Phys. Polon. B}{31}{2000}{2127}
%%CITATION = APPOA,B31,2127;%%

\bibitem{Brueckner:1951} 
\refer{K.~Brueckner}{Phys. Rev.}{82}{1951}{598};
%%CITATION = PHRVA,82,598;%%
% \bibitem{Watson:1951}
\refer{K.M.~Watson,  K.~Brueckner}{Phys. Rev.}{83}{1951}{1};
%%CITATION = PHRVA,83,1;%%
% \bibitem{Watson:1952ji}
%``The Effect Of Final State Interactions On Reaction Cross-Sections,''
\refer{K.M.~Watson}{Phys. Rev.}{88}{1952}{1163};
%%CITATION = PHRVA,88,1163;%%
% \bibitem{GellMann:1953}
\refer{M.~Gell-Mann, M.L.~Goldberger}{Phys. Rev.}{91}{1953}{398};
%%CITATION = PHRVA,91,398;%%
% \bibitem{Migdal:1955} 
\refer{A.B.~Migdal}{Sov. Phys. JETP}{1}{1955}{2}
[ \refer{A.B.~Migdal}{Zh. Eksp. Teor. Fiz.}{28}{1955}{3}];
%%CITATION = SPHJA,1,2;%%
% \bibitem{Fermi:1955} 
\refer{E.~Fermi}{Nuovo Cim. Suppl.}{Vol. II (Serie X) No. 1}{1955}{17}
%%CITATION = NUCUA,2,17;%%

\bibitem{Aitchison:1969tq}
%``Watson's Theorem When There Are Three Strongly Interacting Particles In The
%Final State,''
\refer{I.J.R.~Aitchison, C.~Kacser}{Phys. Rev.}{173}{1968}{1700}
%%CITATION = PHRVA,173,1700;%%

\bibitem{Gordon:1928}
\refer{W.~Gordon}{Z. Phys.}{48}{1928}{180}; 
A.~Sommerfeld: {\em Atombau und Spectrallinien}, F.\ Vieweg und Sohn, Braunschweig, 1939, volume II;
% \bibitem{Guth:1951}
%``Momentum Representation of the Coulomb Scattering Wave Functions,''
\refer{E.~Guth, C.J.~Mullin}{Phys. Rev.}{83}{1951}{667};
%%CITATION = PHRVA,83,667;%%
% \bibitem{Nordsieck:1954}
%``Reduction of an Integral in the Theory of Bremsstrahlung,''
\refer{A.~Nordsieck}{Phys. Rev.}{93}{1954}{785};
%%CITATION = PHRVA,93,785;%%
% \bibitem{Wichmann:1961}
%``Integral Representation for the Nonrelativistic Coulomb Green's Function,''
\refer{E.H.~Wichmann, Ching-Hung Woo}{J. Math. Phys.}{2}{1961}{178};
%%CITATION = JMAPA,2,178;%%
% \bibitem{Mano:1964}
%``Representations for Nonrelativistic Coulomb Green's Function,''
\refer{K.~Mano}{J. Math. Phys.}{5}{1964}{505};
%%CITATION = JMAPA,5,505;%%
% \bibitem{Hostler:1964}
%``Coulomb Green's Functions and the Furry Approximation,''
\refer{L.~Hostler}{J. Math. Phys.}{5}{1964}{591};
%%CITATION = JMAPA,5,591;%%
% \bibitem{Dollard:1964}
%``Asymptotic Convergence and the Coulomb Interaction,''
\refer{J.D.~Dollard}{J. Math. Phys.}{5}{1964}{729}
%%CITATION = JMAPA,5,729;%%

\bibitem{VanHaeringen:1985tp}
H.~van Haeringen: {\it Charged Particle Interactions}, Coulomb Press Leyden, Leiden, 1985
%\href{http://www.slac.stanford.edu/spires/find/hep/www?irn=1784064}{SPIRES entry}

\bibitem{Ahmed:2003}
Z.~Ahmed:
%``A note on Coulomb scattering amplitude,''
quant-ph/0310019;
%%CITATION = QUANT-PH 0310019;%%
% \bibitem{Mukhamedzhanov:2006cs}
%``Completeness of the Coulomb scattering wave functions,''
A.M.~Mukhamedzhanov, M.~Akin:
nucl-th/0602006
%%CITATION = NUCL-TH 0602006;%%

\bibitem{Okubo:1960b}
%``Some Aspects of the Covariant Two-Body Problem. II. The Scattering Problem,''
\refer{S.~Okubo, D.~Feldman}{Phys. Rev.}{117}{1960}{292};
%%CITATION = PHRVA,117,292;%%
% \bibitem{Schwinger:1964}
%``Coulomb Green's Function,''
\refer{J.~Schwinger}{J. Math. Phys.}{5}{1964}{1606};
%%CITATION = JMAPA,5,1606;%%
% \bibitem{Chen:1972}
% ``Remarks on the Construction of the Nonrelativistic Coulomb Green's Function,''
\refer{A.C.~Chen, J.C.Y.~Chen, A.K.~Rajagopal}{Phys. Rev. A}{5}{1972}{2686}
%%CITATION = PHRVA,A5,2686;%%

\bibitem{Ford:1964}
%``Anomalous Behavior of the Coulomb T Matrix,''
\refer{W.F.~Ford}{Phys. Rev.}{133}{1964}{B1616}
%%CITATION = PHRVA,133,B1616;%%

\bibitem{Kowalski:1965}
\refer{K.L.~Kowalski}{Phys. Rev. Lett.}{15}{1965}{798};
%%CITATION = PRLTA,15,798;%%
\refer{H.P.~Noyes}{Phys. Rev. Lett.}{15}{1965}{538}
%%CITATION = PRLTA,15,538;%%

\bibitem{Kok:1982}
%``Half-shell scattering by a screened Coulomb potential,''
\refer{L.P.~Kok, J.W.~de~Maag, R.R.~Bontekoe}{Phys. Rev. C}{26}{1982}{819}
%%CITATION = PHRVA,C26,819;%%

\bibitem{Talukdar:1984}
% ``Coulomb half-shell t matrix,''
\refer{B.~Talukdar, D.K.~Ghosh, T.~Sasakawa}{Phys. Rev. A}{29}{1984}{1865}
%%CITATION = PHRVA,A29,1865;%%

\bibitem{Katsogiannis:1994}
%``Coulomb correction calculations of pp bremsstrahlung,''
\refer{A.~Katsogiannis, K.~Amos, M.~Jetter, H.V.~von~Geramb}{Phys. Rev. C}{49}{1994}{2342}
%%CITATION = PHRVA,C49,2342;%%

\bibitem{Chen:1971}
% ``Off-Shell and On-Shell Unitarity Relations in Two-Body Coulomb Scattering,''
\refer{A.C.~Chen, J.C.Y.~Chen}{Phys. Rev. A}{4}{1971}{2226}
%%CITATION = PHRVA,A4,2226;%%

\bibitem{Kok:1981}
%``Importance of Coulomb Effects in Half-Shell Scattering,''
\refer{L.P.~Kok, H.~van~Haeringen}{Phys. Rev. Lett.}{46}{1981}{1257}
%%CITATION = PRLTA,46,1257;%%

\bibitem{Alston:1988}
% ``Limiting behaviors of off-shell scattering wave functions and T matrices for centrally modified coulomb potentials,''
\refer{S.~Alston}{Phys. Rev. A}{38}{1988}{636}
%%CITATION = PHRVA,A38,636;%%

\bibitem{Dolinskii:1966}
%``Coulomb Effects In Direct Nuclear Reactions,''
\refer{\'{E}.I.~Kolinski\u{\i}, A.M.~Mukhamedzhanov}{Sov. J. Nucl. Phys.}{3}{1966}{180} [{\em J. Nucl. Phys. (U.S.S.R.)} {\bf 3} (1966) 252];
%%CITATION = SJNCA,3,180;%%
% \bibitem{Vincent:1974}
  %``Accurate momentum-space method for scattering by nuclear and Coulomb potentials,''
\refer{C.M.~Vincent, S.C.~Phatak}{Phys. Rev. C}{10}{1974}{391}
%%CITATION = PHRVA,C10,391;%%

\bibitem{Bajzer:1987}
\refer{\v{Z}.~Bajzer}{Few Body Systems}{2}{1987}{9};
%%CITATION = FBSYE,2,9;%%
% \bibitem{Vlahovic:1994}
%``Effect of Coulomb interacton in quasifree scattering and quasifree reactions in three body breakup processes,''
\refer{B.~Vlahovi\'{c} {\it et al.}}{Phys. Rev. C}{49}{1994}{2643}
%%CITATION = PHRVA,C49,2643;%%

\bibitem{Moalem:1995} 
A.\ Moalem, L.\ Razdolskaya, E.\ Gedalin: hep-ph/9505264.
%%CITATION = HEP-PH 9505264;%%

\bibitem{Alt:2004fi}
%``Coulomb Fourier transformation: A novel approach to three-body scattering
%with charged particles,''
\refer{E.O.~Alt, S.B.~Levin, S.L.~Yakovlev}{Phys. Rev. C}{69}{2004}{034002}
%%CITATION = PHRVA,C69,034002;%%

\bibitem{Mukhamedzhanov:2005xc}
%``The leading asymptotic terms of the three-body Coulomb scattering wave
%function,''
A.M.~Mukhamedzhanov, A.S.~Kadyrov, F.~Pirlepesov: nucl-th/0509012
%%CITATION = NUCL-TH 0509012;%%

\bibitem{Alt:1978yd}
%``Coulomb Effects In Three-Body Reactions With Two Charged Particles,''
\refer{E.O.~Alt {\it et al.}}{Phys. Rev. C}{17}{1978}{1981}; 
%%CITATION = PHRVA,C17,1981;%%
E.O.\ Alt, W. Sandhas: pp.\ 1-95 in {\it Coulomb Interactions in Nuclear and Atomic Few-Body Collisions}, eds. F.S.~Levin, D.~Micha, Plenum, New York, 1996

\bibitem{Dollard:1968}
%``Screening in the Schroedinger Theory of Scattering,''
\refer{J.D.~Dollard}{J. Math. Phys.}{9}{1968}{620};
%%CITATION = JMAPA,9,620;%%
% \bibitem{Semon:1974pj}
%``Scattering By Potentials With Coulomb Tails,''
\refer{M.D.~Semon, J.R.~Taylor}{Nuovo Cim. A}{26}{1975}{48}
%%CITATION = NUCIA,A26,48;%%

\bibitem{Deltuva:2005wx}
%``Momentum-space treatment of Coulomb interaction in three-nucleon reactions
%with two protons,''
\refer{A.~Deltuva {\it et al.}}{Phys. Rev. C}{71}{2005}{054005},
%%CITATION = NUCL-TH 0503012;%%
% \bibitem{Deltuva:2005cc}
%``Momentum-space description of three-nucleon breakup reactions including the
%Coulomb interaction,''
% A.~Deltuva, A.~C.~Fonseca and P.~U.~Sauer,
{\bf 72} (2005) 054004
[Erratum {\bf 72} (2005) 059903]
%%CITATION = NUCL-TH 0509034;%%

\bibitem{Faldt:1997jm}
\refer{G.F\"{a}ldt, C.~Wilkin}{Phys. Rev. C}{56}{1997}{2067};
%%CITATION = NUCL-TH 9704056;%%
% \bibitem{Kleefeld:1999nr}
%``Threshold behaviour of meson nucleon S11(1535) vertex functions and
%determination of the S11(1535) mixing angle,''
\refer{F.~Kleefeld}{$\pi$N Newslett.}{15}{1999}{250}
[nucl-th/9910010];
%%CITATION = NUCL-TH 9910010;%%
% \bibitem{Kleefeld:2000as}
%``ISI / FSI in threshold meson production: Onshell approach and Coulomb
%problem,''
\refer{F.~Kleefeld}{Acta Phys. Polon. B}{31}{2000}{2225};
%%CITATION = NUCL-TH 0005037;%%
% \bibitem{Kleefeld:2000vm}
\refer{F.~Kleefeld}{Nucl. Phys. A}{689}{2001}{471};
%%CITATION = NUCL-TH 0010002;%%
% \bibitem{Hanhart:1998rn}
%``On the treatment of N N interaction effects in meson production in N N
%collisions,''
\refer{C.~Hanhart, K.~Nakayama}{Phys. Lett. B}{454}{1999}{176}
%%CITATION = NUCL-TH 9809059;%%

\bibitem{Deloff:2003te}
%``Phenomenology of p p $\to$ p p eta reaction close to threshold,''
\refer{A.~Deloff}{Phys. Rev. C}{69}{2004}{035206};
%%CITATION = NUCL-TH 0309059;%%
% \bibitem{Deloff:2004da}
\refer{A.~Deloff}{Schriften des FZ J\"ulich, Series Matter and Material}{21}{2004}{206} [nucl-th/0406069]
%%CITATION = NUCL-TH 0406069;%%

\bibitem{Baryshevskii:2004}
% Regularization of the Coulomb scattering problem
\refer{V.G.~Baryshevskii, I.D.~Feranchuk, P.B.~Kats}{Phys. Rev. A}{70}{2004}{052701} [quant-ph/040305]
%%CITATION = PHRVA,A70,052701;%%

\bibitem{Pineda:2004mx}
%``The chiral structure of the Lamb shift and the definition of the proton
%radius,''
\refer{A.~Pineda}{Phys. Rev. C}{71}{2005}{065205}
%%CITATION = HEP-PH 0412142;%%
( see also 
% \bibitem{Sick:2003gm}
%``On the rms-radius of the proton,''
\refer{I.~Sick}{Phys. Lett. B}{576}{2003}{62}
%%CITATION = NUCL-EX 0310008;%%
)

\bibitem{Klaja:2005a} P.\ Klaja {\it et al.}: {\em Correlation femtoscopy for studying $\eta$ meson production mechanism}, submitted for publication in {\em Acta Physica Slovaca}

\bibitem{Gasparyan:2005fk}
%``Extraction of scattering lengths from final-state interactions,''
\refer{A.~Gasparyan, J.~Haidenbauer, C.~Hanhart}{Phys. Rev. C}{72}{2005}{034006}
%%CITATION = NUCL-TH 0506067;%%

\bibitem{Kadyrov:2005xm}
%``Scattering theory for arbitrary potentials,''
\refer{A.S.~Kadyrov, I.~Bray, A.M.~Mukhamedzhanov, A.T.~Stelbovics}{Phys. Rev. A}{72}{2005}{032712}
%%CITATION = NUCL-TH 0508014;%%

%%%%%%%%%%%%%%%%%%%% LSM %%%%%%%%%%%%%%%%%%%%%%%%%%%%%%%%%%%%%%

\bibitem{Levy:1967a}
\refer{M. L\'{e}vy}{Nuovo Cim.}{LII A}{1967}{23};
%%CITATION = NUCIA,52A,23;%%
% \bibitem{Gasiorowicz:1969kn}
\refer{S.~Gasiorowicz, D.A.~Geffen}{Rev. Mod. Phys.}{41}{1969}{531};
%%CITATION = RMPHA,41,531;%%
% \bibitem{Cabibbo:1970uc}
\refer{N.~Cabibbo, L.~Maiani}{Phys. Rev. D}{1}{1970}{707};
%%CITATION = PHRVA,D1,707;%%
% \bibitem{Schechter:1993tc}
\refer{J.~Schechter, Y.~Ueda}{Phys. Rev. D}{3}{1971}{2874} [Erratum {\bf 8} (1973) 987];
%%CITATION = PHRVA,D3,2874;%%
% \bibitem{Chan:1973qn}
\refer{L.H.~Chan, R.W.~Haymaker}{Phys. Rev. D}{7}{1973}{402, 415};
%%CITATION = PHRVA,D7,402;%%
%\bibitem{Chan:1973qi}
%%CITATION = PHRVA,D7,415;%%
% \bibitem{Geddes:1979im}
\refer{H.B.~Geddes, R.H.~Graham}{Phys. Rev. D}{21}{1980}{749};
%%CITATION = PHRVA,D21,749;%%
% \bibitem{Napsuciale:1998ip}
M.~Napsuciale:
hep-ph/9803396;
%%CITATION = HEP-PH 9803396;%%
%\bibitem{Napsuciale:2001sm}
\refer{M.~Napsuciale, S.~Rodriguez}{Int. J. Mod. Phys. A}{16}{2001}{3011};
%%CITATION = HEP-PH 0204149;%%
% \bibitem{'tHooft:1986nc}
\refer{G.~'t~Hooft}{Phys. Rept.}{142}{1986}{357};
%%CITATION = PRPLC,142,357;%%
% \bibitem{'tHooft:1999jc}
G.~'t Hooft:
hep-th/9903189;
%%CITATION = HEP-TH 9903189;%%
% \bibitem{Tornqvist:1999tn}
\refer{N.A.~T\"{o}rnqvist}{Eur. Phys. J. C}{11}{1999}{359}
%%CITATION = HEP-PH 9905282;%%

\bibitem{Delbourgo:1998kg}
\refer{R.~Delbourgo, M.D.~Scadron}{Int. J. Mod. Phys. A}{13}{1998}{657}
[hep-ph/9807504]
%%CITATION = HEP-PH 9807504;%%

\bibitem{Scadron:2006mq}
M.~D.~Scadron, F.~Kleefeld, G.~Rupp:
%``Pion chiral symmetry breaking in the quark-level linear sigma model and
%chiral perturbation theory,''
hep-ph/0601196
%%CITATION = HEP-PH 0601196;%%

\bibitem{Klabucar:2001gr}
\refer{D.~Kekez, D.~Klabu\v{c}ar, M.D.~Scadron}{J.\ Phys.\ G}{27}{2001}{1775},
%%CITATION = HEP-PH 0101324;%%
% \bibitem{Kekez:2000aw}
{\bf 26} (2000) 1335
%%CITATION = HEP-PH 0003234;%%

\bibitem{Kleefeld:2001ds}
%``Identifying the quark content of the isoscalar scalar mesons f0(980),
%f0(1370), and f0(1500) from weak and electromagnetic processes,''
\refer{F.~Kleefeld, E.~van Beveren, G.~Rupp, M.D.~Scadron}{Phys.\ Rev.\ D}{66}{2002}{034007}
%%CITATION = HEP-PH 0109158;%%

\bibitem{Kroll:2005sd}
P.~Kroll: hep-ph/0509031;
%%CITATION = HEP-PH 0509031;%%
% \bibitem{Muller:2004vf}
\refer{S.E.~M\"uller  [KLOE Coll.]}
{Int. J. Mod. Phys. A}{20}{2005}{1888}
%%CITATION = HEP-EX 0411081;%%

\bibitem{Escribano:2005qq}
\refer{R.~Escribano, J.M.~Frere}{JHEP}{06}{2005}{029}
[hep-ph/0501072];
%%CITATION = HEP-PH 0501072;%%
% \bibitem{Aloisio:2002vm}
\refer{A.~Aloisio {\it et al.} [KLOE Coll.]}{Phys. Lett. B}{541}{2002}{45};
%%CITATION = HEP-EX 0206010;%%
% \bibitem{Bramon:1997va}
\refer{A.~Bramon, R.~Escribano, M.D.~Scadron}{Eur. Phys. J. C}{7}{1999}{271}
%%CITATION = HEP-PH 9711229;%%

\bibitem{Bramon:1997mf}
\refer{A.~Bramon, R.~Escribano, M.D.~Scadron}{Phys. Lett. B}{403}{1997}{339},
%%CITATION = HEP-PH 9703313;%%
{\bf 503} (2001) 271;
%%CITATION = HEP-PH 0012049;%%
% \bibitem{Amsler:1997up}
C.~Amsler: {\it Rev. Mod. Phys.} {\bf 70} (1998) 1293;
%%CITATION = HEP-EX 9708025;%%
% \bibitem{Amsler:1992wm}
\refer{C.~Amsler {\it et al.} [Crystal Barrel Coll.]}{Phys. Lett. B}{294}{1992}{451}
%%CITATION = PHLTA,B294,451;%%

\bibitem{Truong:1991gv}
\refer{T.N.~Truong}{Phys. Rev. Lett.}{67}{1991}{2260};
%%CITATION = PRLTA,67,2260;%%
% \bibitem{Kleefeld:2005xx}
F.~Kleefeld: PoS(HEP2005)108
%``On a new unitarization scheme inspired by Dalitz and Tuan applied to meson
%meson and meson baryon scattering,''
[hep-ph/0511096]
%%CITATION = HEP-PH 0511096;%%


\bibitem{Kleefeld:2003bw}
%``The light and heavy scalars in unitarized coupled channel and  Lagrangian
%approaches,''
\refer{F.~Kleefeld}{AIP Conf. Proc.}{717}{2004}{332};
%%CITATION = HEP-PH 0310320;%%
% \bibitem{Rupp:2004rf}
%``Scalar mesons and Adler zeros,''
\refer{G.~Rupp {\it et al.}}{AIP Conf. Proc.}{756}{2005}{360}
%%CITATION = HEP-PH 0412078;%%

\bibitem{vanBeveren:2002gy}
\refer{E.~van Beveren {\it et al.}}{AIP Conf. Proc.}{660}{2003}{353};
%%CITATION = HEP-PH 0211411;%%
% \bibitem{vanBeveren:2005ha}
%``From the Kappa via the Ds0*(2317) to the chi_c0: connecting light and heavy
%scalar mesons,''
E.~van Beveren {\it et al.}: hep-ph/0509351;
%%CITATION = HEP-PH 0509351;%%
% \bibitem{vanBeveren:2003kd}
\refer{E.~van Beveren, G.~Rupp}{Phys. Rev. Lett.}{91}{2003}{012003};
%%CITATION = HEP-PH 0305035;%%
% \bibitem{Tornqvist:1995kr}
\refer{N.A.~T\"{o}rnqvist}{Z. Phys. C}{68}{1995}{647}
%%CITATION = HEP-PH 9504372;%%

\bibitem{vanBeveren:2001kf}
%``Modified Breit-Wigner formula for mesonic resonances describing OZI  decays of confined q anti-q states and the light scalar mesons,''
\refer{E.~van Beveren, G.~Rupp}{Eur. Phys. J. C}{22}{2001}{493}
%%CITATION = HEP-EX 0106077;%%

\bibitem{VanBeveren:1986ea}
\refer{E.~van Beveren {\it et al.}}{Z. Phys. C}{30}{1986}{615}
%%CITATION = ZEPYA,C30,615;%%

\bibitem{ruppthesis1} G.\ Rupp: doctoral thesis, Catholic Univ.\ of Nijmegen, 1982;
% \bibitem{vanBeveren:bd}
% E.~van Beveren, C.~Dullemond, G.~Rupp,
%``Spectrum And Strong Decays Of Charmonium,''
\refer{E.~van Beveren et al.}{Phys. Rev. D}{21}{1980}{772}, 
%%CITATION = PHRVA,D21,772;%%
{\bf 22} (1980) 787,
%%CITATION = PHRVA,D22,787;%%
% E.~van Beveren, G.~Rupp, T.~A.~Rijken, C.~Dullemond,
%``Radial Spectra And Hadronic Decay Widths Of Light And Heavy Mesons,''
{\bf 27} (1983) 1527;
%%CITATION = PHRVA,D27,1527;%%
%``Matrix Elements Of The Exchange Operator For Arbitrary Angular Momentum Two Meson States,''
\refer{J.E.~Ribeiro}{Phys. Rev. D}{25}{1982}{2406}
%%CITATION = PHRVA,D25,2406;%%

\bibitem{Aston:1987ir}
%``A Study Of K- Pi+ Scattering In The Reaction K- P $\to$ K- Pi+ N At
%11-Gev/C,''
\refer{D.~Aston {\it et al.}}{Nucl. Phys. B}{296}{1988}{493}
%%CITATION = NUPHA,B296,493;%%

\bibitem{Bugg:2003kj}
\refer{D.V.~Bugg}{Phys. Lett. B}{572}{2003}{1} [Erratum {\bf 595} (2004) 556]; 
%%CITATION = PHLTA,B572,1;%%
% \bibitem{Bugg:2004xu}
{\it Phys. Rept.} {\bf 397} (2004) 257
%%CITATION = HEP-EX 0412045;%%

\bibitem{Kleefeld:2005hd}
\refer{F.~Kleefeld}{Czech J. Phys.}{55}{2005}{1123}
%``On (non-Hermitian) Lagrangeans in (particle) physics and their dynamical
%generation,''
[hep-th/0506140]
%%CITATION = HEP-TH 0506140;%%
\end{thebibliography}
\end{document}